\documentclass[%
    reprint,
    showpacs,
    amsmath,amssymb,
    aps,
    prb,
    floatfix,
]{revtex4-1}

\usepackage{graphicx}
\usepackage{dcolumn}
\usepackage{bm}
\usepackage{braket}
\usepackage{subfigure}
\usepackage{hyperref}
\hypersetup{
   colorlinks=true,
      linkcolor=blue,
      citecolor=blue,
        urlcolor=blue,
}
\begin{document}
\title{Site-dependent magnetism of Ni adatoms on MgO/Ag(001)}
\author{Oliver R. Albertini}
    \email{ora@georgetown.edu}
    \affiliation{
        Department of Physics, Georgetown University, Washington, DC 20057, USA 
    }
\author{Amy Y. Liu}
    \affiliation{
        Department of Physics, Georgetown University, Washington, DC 20057, USA 
    }
\author{Barbara A. Jones}
    \affiliation{
        IBM Almaden Research Center, San Jose, CA 95120, USA
    }
\date{\today}

\begin{abstract}
We examine the adsorption of a single Ni atom on a monolayer of MgO on a Ag substrate using DFT and DFT$+$$U$ computational approaches. 
We find that the electronic and magnetic properties vary considerably across the three binding sites of the surface.
Two of the binding sites are competitive in energy, and the preferred site depends on the strength of the on-site Coulomb interaction $U$. 
These results can be understood in terms of the competition between bonding and magnetism for surface adsorbed transition metal atoms. 
Comparisons are made with a recent experimental and theoretical study of Co on MgO/Ag, and implications for scanning tunneling microscopy experiments on the Ni system are discussed.
\end{abstract}

                        
\maketitle

\section{\label{sec:intro}Introduction}

The study of magnetic adatoms on surfaces has drawn recent attention due to 
possible applications in the realm of magnetic storage and quantum computation. 
The density of magnetic storage  has enjoyed exponential growth 
for several decades, but this growth will eventually slow as particle 
sizes approach the superparamagnetic regime.\cite{neel} An alternative,
bottom-up approach is to start from the atomic limit.  With the scanning 
tunneling microscope (STM),  single atoms can be moved around on a surface to 
construct desired nanostructures, and the STM can also be used to probe 
the electronic and magnetic properties of those nanostructures.
Such experiments have found, for example, that a single Fe atom on 
a Cu$_2$N monolayer (ML) on Cu has a large magnetic anisotropy energy,\cite{hirjibehedin}
and that a magnetic bit consisting of an array of 12 Fe atoms on the same surface 
has a stable moment  that can stay in the `on' or `off' state for
hours at cryogenic temperatures ($\sim$ 1 K).\cite{loth} 

A recent STM study of a single Co atom on a ML of MgO on the Ag(001) 
surface found that the Co atom 
maintains its gas phase spin of $S = 3/2$, and, because of the axial 
properties of the ligand field, it also maintains its orbital moment on the 
surface ($L = 3$). The resulting  magnetic anisotropy is the 
largest possible for a $3d$ transition metal, set by the spin-orbit splitting 
and  orbital angular momentum. The measured spin relaxation time of
200 $\mu$s is three orders of magnitude larger than typical for a
transition-metal atom on an insulating substrate.\cite{cobalt} 

Here we investigate the structural, electronic, and magnetic properties of Ni 
adatoms on the same substrate, MgO/Ag(001), using density functional methods. 
Previous authors \cite{yudanov, neyman1997, lopez, markovits2001, 
markovits2003, neyman2004, fernandez, dong} have studied the adsorption 
of transition metal atoms on an MgO substrate using ab-initio 
techniques.
These studies, which employed various surface models and approximations
for the exchange-correlation functional, are in general agreement that the 
preferred binding site of a Ni adatom on the MgO(001) surface is on top 
of an O atom, and that structural distortion of the MgO surface upon Ni 
adsorption is minimal. The situation is less clear when it comes to the
spin state of the adatom, since the $s$-$d$ transition energy is so small in 
Ni. Calculations that employ the generalized gradient approximation for the 
exchange-correlation functional generally predict a full quenching of the 
Ni moment on MgO, while partial inclusion of Fock exchange as in the 
B3LYP hybrid functional predicts that the triplet spin state is slightly more 
favorable.\cite{markovits2001, markovits2003, lopez}
\\ \indent The system studied in the current work differs from previous studies of 
Ni adsorption on MgO because the substrate 
consists of a single layer of MgO atop the Ag(001) surface.  This
aligns more closely with STM experiments that  use a 
thin insulating layer to decouple the magnetic adatom
from the conducting substrate that is necessary for electrically probing the
system.  We show that, in contrast to the surface layer of an 
MgO substrate, the MgO ML on Ag can deform significantly due to
interactions with the Ni, resulting in a very different potential energy 
surface for adsorption.
Further, we investigate how the interaction of Ni with the MgO/Ag substrate is 
affected by on-site Coulomb interactions and find that the preferred
binding site depends on the interaction strength $U$.   Unlike 
the case of Co adatoms on the MgO/Ag substrate, the Ni spin moment is 
always lower than that of the isolated atom. The degree to
which the moment is reduced depends strongly on the binding site.  These results
can be understood by considering the Ni $3d$-$4s$ hybrid orbitals that 
participate in bonding at different sites.  We conclude with a discussion
about experiments that could corroborate our  findings. 
\section{\label{sec:method}Method}

\subsection{\label{sec:comp_methods}Computational methods}

Two sets of density-functional-theory (DFT) calculations were carried out, 
one using the linearized augmented plane wave (LAPW) method  as implemented in 
WIEN2K,\cite{wien2k} and the other using the projector augmented wave (PAW) method\cite{PAW} in the 
VASP package.\cite{vasp1} Both started with the Perdew, Burke and  Ernzerhof  formulation
of the generalized gradient approximation (GGA) for the exchange-correlation 
functional.\cite{pbe}  In the LAPW calculations, structures 
for each binding site were relaxed within the GGA. These structures were then 
used to calculate Hubbard $U$ values for each site using the constrained DFT 
method of Madsen and Nov\'{a}k.\cite{madsen-novak}  GGA$+$$U$ 
calculations were then carried out using those site-specific values of $U$ 
to examine electronic and magnetic properties, without further structural 
relaxation.  On the other hand, PAW calculations were used to optimize 
structures within both GGA and GGA$+$$U$. However, since the total energy in 
DFT$+$$U$ methods depends on the value of the on-site Coulomb interaction strength, 
the same value of $U$ was used for each adsorption site to allow comparison 
of binding energies. In all GGA$+$$U$ calculations, a rotationally invariant method in which only the difference $U-J$ is meaningful was employed.\cite{dudarev}

Within GGA, the two sets of calculations yielded very similar 
results for adsorption geometries, binding energies, and electronic and 
magnetic structure, as expected. Although the calculations were used for 
different purposes in assessing the effect of on-site Coulomb repulsion, 
the GGA$+$$U$ results from the two methods were generally consistent and
displayed the same trends.\footnote{All-electron LAPW calculations used $R_{mt}K_{max} =  7$, where $R_{mt}$ is the smallest muffin-tin radius in the unit cell, for all calculations. Structural relaxations were carried out with k-point meshes of $6\times6\times1$ and finite temperature smearing of 0.001 Ry. Denser k-point grids of up to $19\times19\times1$ were used for more accurate magnetic moments and densities of states. VASP calculations were carried out using a plane-wave cutoff of 500 eV, k-point sampling of $8\times8\times1$  grids, and a Gaussian smearing width of 0.02 eV.}

\subsection{\label{sec:geometry}Supercell geometry and binding sites}

STM experiments require a conducting substrate, yet for magnetic
nanostructures, it is desirable to suppress interaction between
the adatoms  and a metallic substrate. Hence the adatom is
often placed on a thin insulating layer at the surface of the substrate. It has been shown experimentally that ultrathin ionic insulating layers can shield the adatom from interaction with the surface.\cite{repp} 
In a recent experimental study of Co adatoms, a single atomic layer of MgO was
used as the insulating layer above an Ag substrate.\cite{susanne,cobalt} 
The lattice constants of MgO and Ag are well matched (4.19 \AA~ and 4.09 \AA,  
respectively), and DFT calculations of a single layer of MgO on the Ag(100)
surface have found that is it energetically favorable for the
O atoms in the MgO layer to sit above the Ag atoms.\cite{cobalt}  

Here we used this alignment of MgO on Ag(001) in inversion 
symmetric (001) slabs of at least five Ag layers sandwiched between 
MLs of MgO.  We found only minor differences in results for 
slabs containing five versus seven Ag layers.  The in-plane lattice constant 
was fixed at the bulk Ag value,  and an isolated Ni adatom
on the MgO/Ag surface was modeled using a $3/\sqrt{2}\times3/\sqrt{2}$ 
supercell of the slab with one Ni atom on each surface, corresponding
to a lateral separation of $8.68$ \AA~ between adatoms. 
In the out-of-plane direction, the supercells contained  7 to 8 
layers of vacuum.

\begin{figure}[t]
    \subfigure[~O site]{\includegraphics[height=2.8cm]{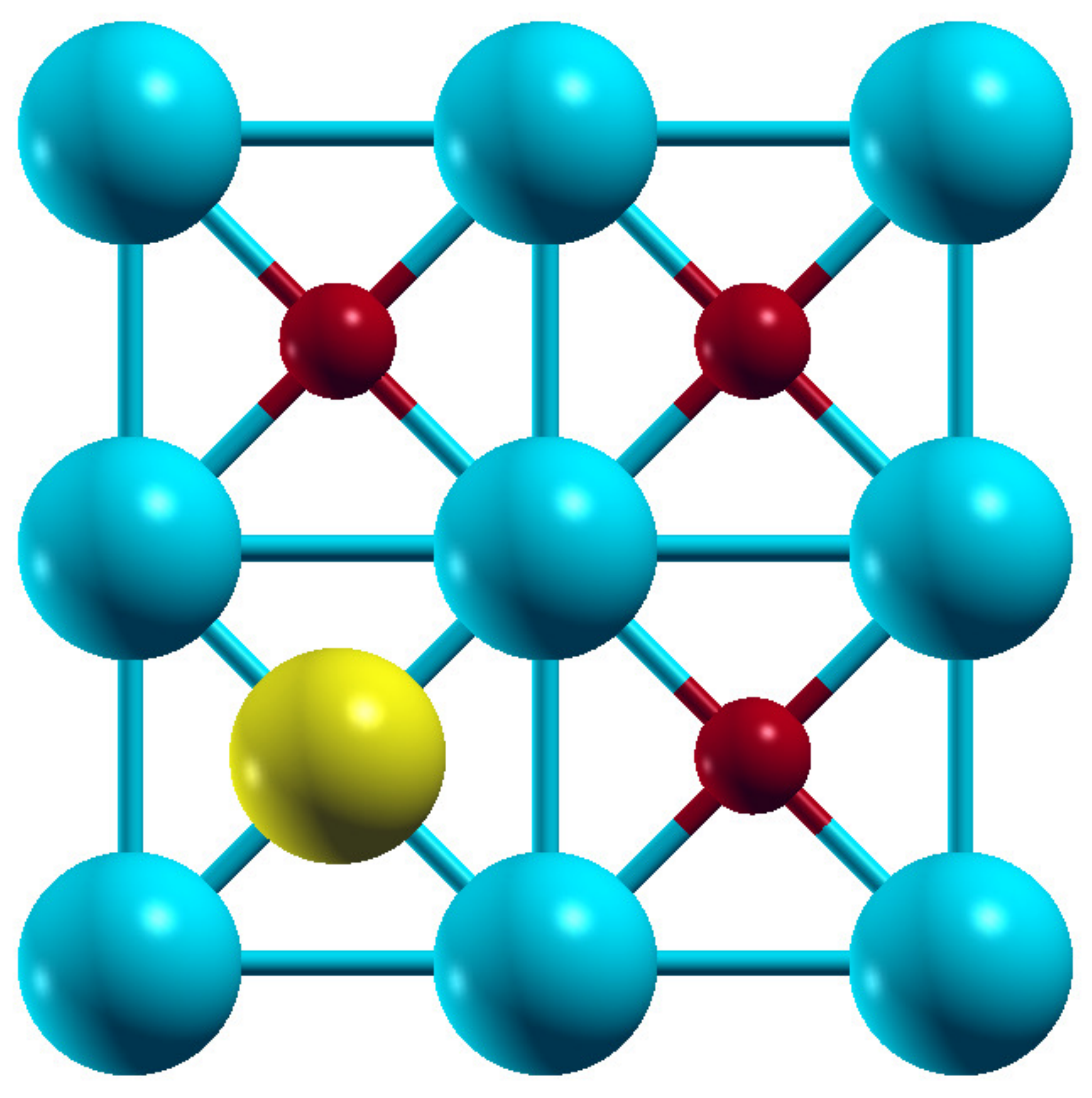}}
    \subfigure[~Mg site]{\includegraphics[height=2.8cm]{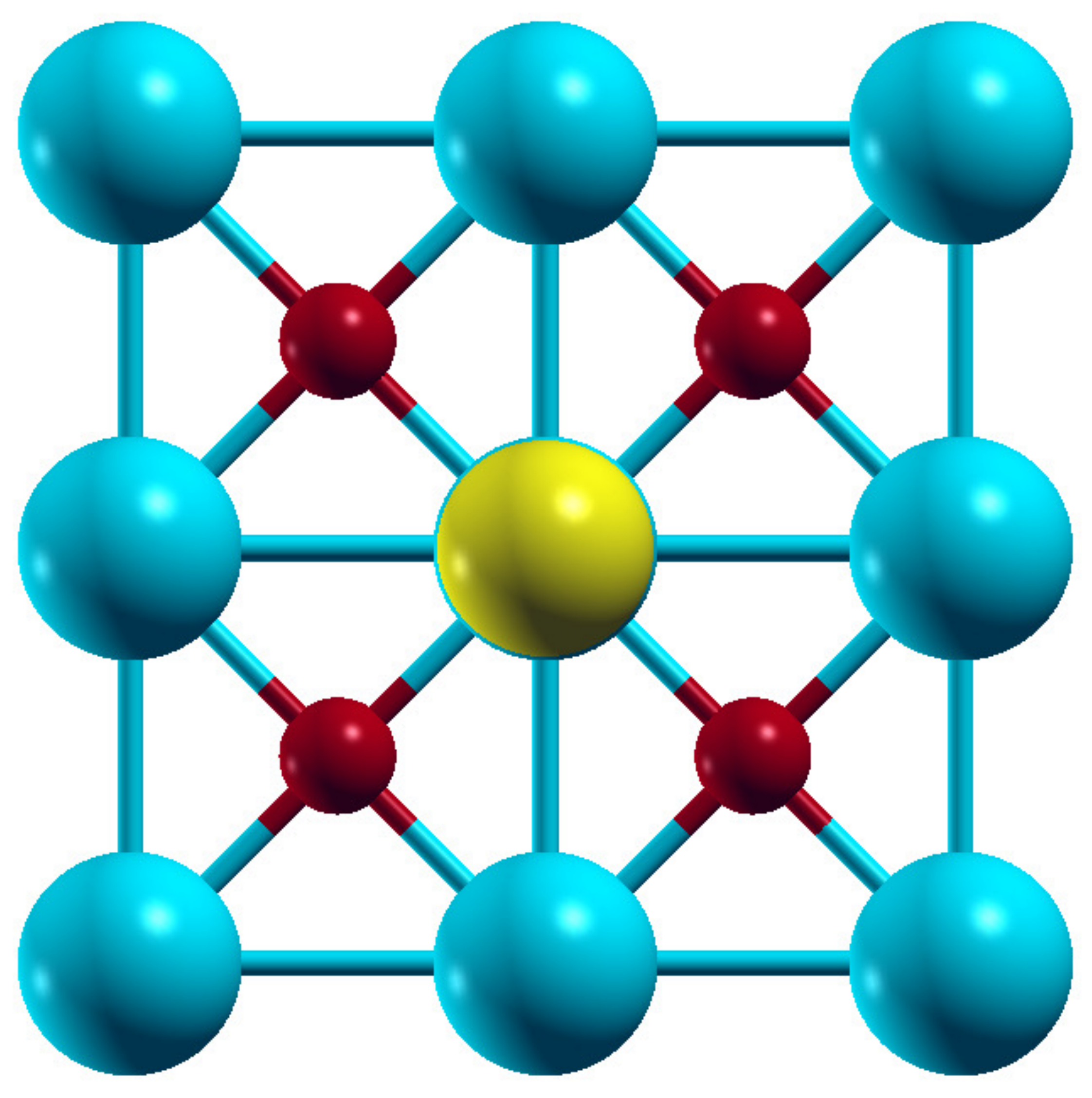}}
    \subfigure[~hollow site]{\includegraphics[height=2.8cm]{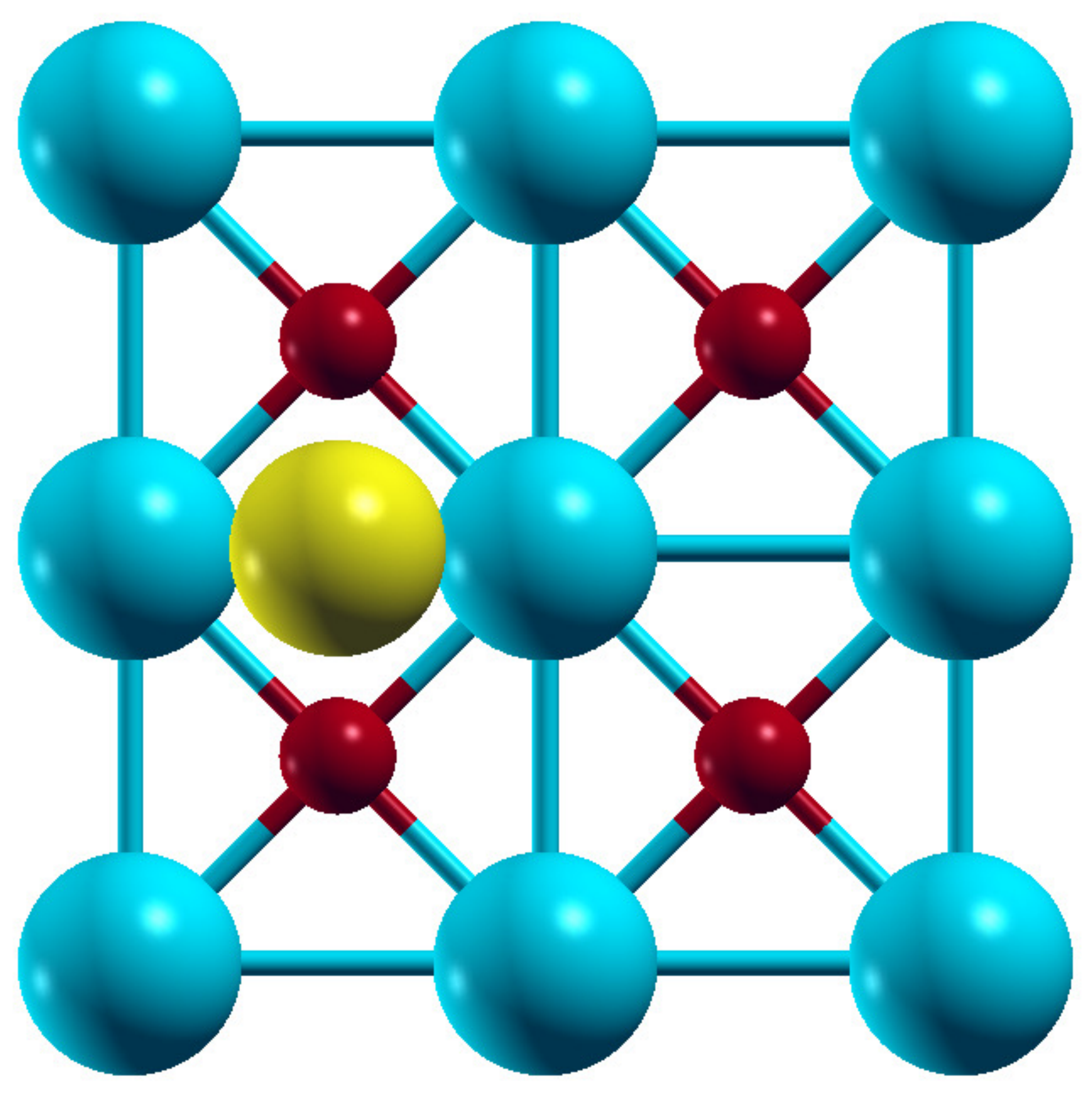}}
    \caption{(Color online) Top view of the three binding sites. Cyan spheres represent Mg, small red spheres O, and the yellow spheres Ni.}
\label{fig:struct}
\end{figure}

Three high-symmetry binding sites on the MgO surface were 
considered, as shown in Fig. \ref{fig:struct}: Ni on top of an O atom, 
Ni on top of an Mg atom, and Ni above the center of the square formed by 
nearest-neighbor Mg and O sites. These will be referred to as the O site, 
the Mg site, and the hollow site, respectively. 

\section{\label{sec:results}Results \& Discussion}
\subsection{\label{sec:energies} Binding energetics and geometries}

\begin{figure}[t]
    \begin{center}
        \includegraphics[width=0.85\linewidth]{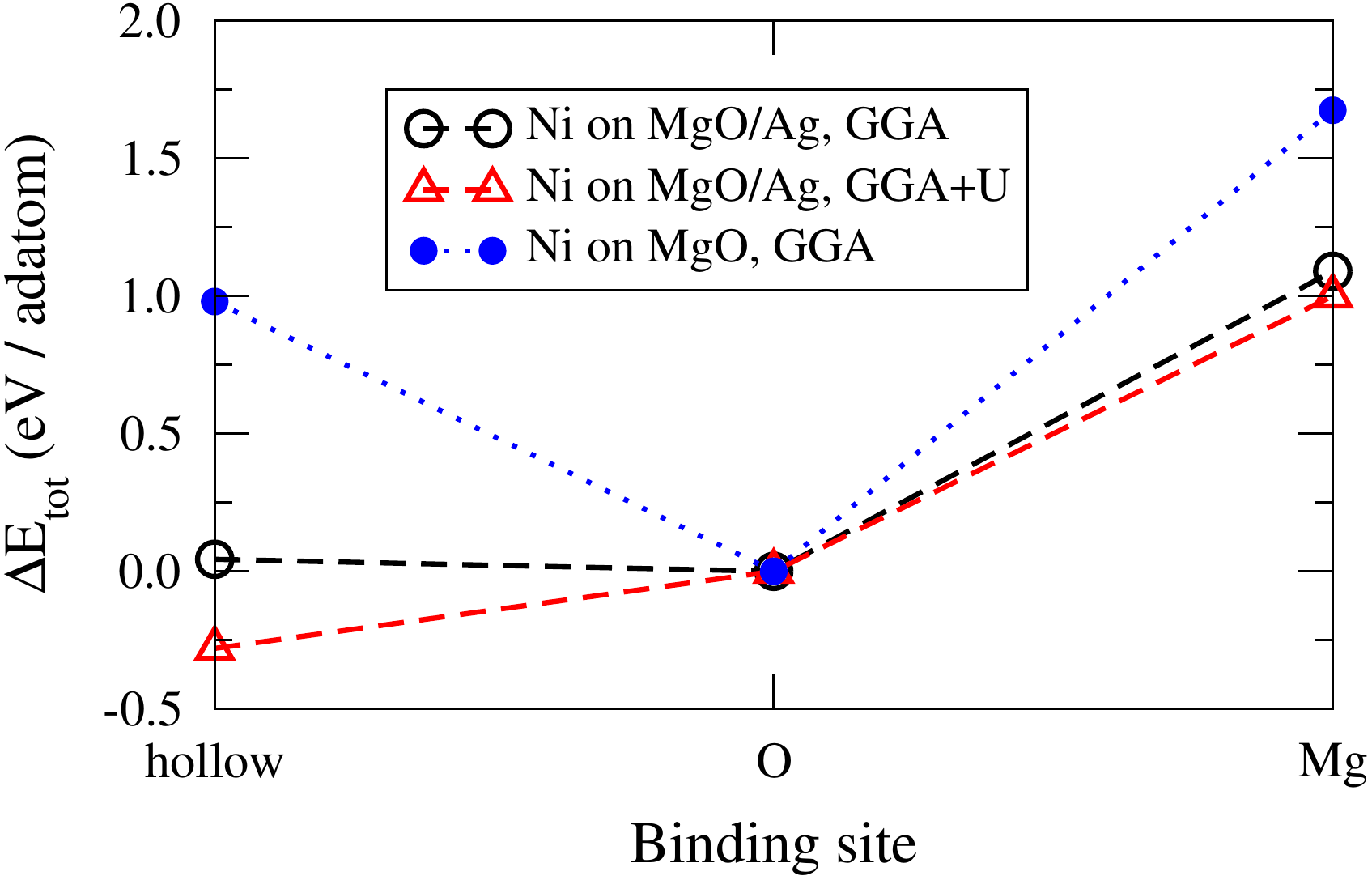}
    \end{center}
    \caption{(Color online) Total energy per Ni adatom 
        on MgO/Ag, calculated for structures relaxed within GGA and 
        GGA$+$$U$ ($U = 4$ eV) from PAW calculations.  For comparison,
        energies calculated for Ni adatoms on MgO are also shown. 
        All energies are plotted relative to the O site energy.}
    \label{fig:energies}
\end{figure}

Since GGA and GGA$+$$U$ do not fully describe important correlation 
effects in the isolated Ni atom, the calculated binding energies
are not as reliable as {\em differences} in binding
energy between different adsorption sites.  Figure \ref{fig:energies}
shows the total energy relative to that of the O site. 
Within GGA, the O site is slightly favored over the hollow site, while
the Mg site is about 1 eV higher in energy.  
For comparison, we also plot our results for 
Ni on an MgO substrate. Similar to previous reports,\cite{dong} 
we find that on the MgO substrate, Ni clearly favors the O site, with 
the hollow and Mg sites lying about 1 and 1.7 eV higher in energy,
respectively.

The difference between the potential energy surfaces 
for Ni adsorption on MgO/Ag versus on MgO can be 
attributed to the greater freedom that the MgO ML has to deform 
to accommodate the adatom. 
The pure MgO substrate remains very flat upon Ni adsorption. When Ni 
is on the O or Mg site, the displacement of surface atoms is 
negligible, and when it is on the hollow site, the neighboring O atoms 
displace out of the plane by about 0.09 \AA.  The situation is very different 
for Ni on MgO/Ag.  Table \ref{tab:structure} lists the 
distance between the adatom and its nearest Mg and O 
neighbors, as well as the vertical displacement $\Delta_z$, of those  
Mg and O atoms.  For each adsorption site,
at least one of the neighboring atoms is displaced vertically by 
more than 0.1 \AA.  The deformation of the MgO ML is most 
striking when Ni is on the hollow site, with neighboring O atoms being 
pulled up out of the MgO ML by nearly 0.5 \AA. 
This allows the hollow site to be competitive in energy with the O site on 
MgO/Ag, in contrast to what happens on the MgO surface.

\begin{table}[b]
	\caption{\label{tab:structure} 
		Relaxed distance between adatom and nearest-neighbor Mg and O
        atoms, and height of these atoms above the MgO ML.
        All distances are in \AA. Results were obtained within
        the GGA using the all-electron LAPW method.}
    \begin{ruledtabular}
        \begin{tabular}{cccccc}
            Adatom & Binding site & $d_{\text{ad-O}}$ & $d_{\text{ad-Mg}}$ & $\Delta z_{\text{O}}$ & $\Delta z_{\text{Mg}}$ \\
            \hline \\[-.80em]
            Ni& O site & $1.79$ & $2.87$ & $+0.15$ & $+0.06$ \\
            Ni& hollow site & $1.90$ & $2.53$ & $+0.48$ & $-0.18$ \\
            Ni& Mg site & $3.63$ & $2.73$ & $-0.15$ & $+0.20$ \\
            Co& O site & $1.85$ & $2.95$ & $-0.06$ & $-0.20$ \\
        \end{tabular}
    \end{ruledtabular}
\end{table}

To examine the influence of local correlations on the binding energetics
of Ni on MgO/Ag, we present GGA$+$$U$ results. Since total energies 
can only be compared for calculations that use the same value of
$U$, structures were relaxed assuming $U=4$ eV for all sites.
GGA$+$$U$ predicts that the O and hollow sites reverse order, 
as shown in Fig. \ref{fig:energies}.  The site above an Mg atom 
still remains much less favorable than either the O or the hollow site.    
Of course the Coulomb interaction $U$ depends on the local
environment of the Ni atom, so it is an approximation to assume a 
fixed value for all adsorption sites. Nevertheless, these results
demonstrate that on-site correlations can change the preferred adsorption 
geometry. A similar effect has been reported for the adsorption of
$3d$ transition metals on graphene.\cite{wehling}

For a Ni adatom above an O atom, the Ni-O bond is weakened by
on-site correlations, as evidenced by an increase of about $8\%$ in the 
Ni-O bond length in going from $U=0$ to $U=4$ eV.  
On the hollow and Mg sites, differences 
between geometries relaxed within GGA and GGA$+$$U$ are much smaller. 

Given the limitations of the DFT$+$$U$ method, we are not able to 
definitively predict whether the O site or hollow site is preferred. 
However,  it is likely that the Mg site is uncompetitive, and not experimentally feasible. Hence the remainder 
of the paper deals primarily with the hollow and O binding sites, comparing
their electronic and magnetic properties. 
\subsection{\label{sec:electronic_prop}Electronic and Magnetic Properties}

Based on the geometries relaxed within GGA, our constrained DFT calculations
yield Coulomb parameters $U$ = 4.6, 6.0, and 5.0 eV for Ni $3d$ electrons
on the O, hollow, and Mg sites, respectively.  The GGA$+$$U$ results presented 
in this section use these site-specific values of $U$, but assume the 
GGA-relaxed geometries.  As mentioned, the O-site geometry is somewhat 
sensitive to the inclusion of on-site Coulomb  repulsion, but the Mg-site 
and hollow site geometries are not. 

\begin{table}[b]
	\caption{\label{tab:magmom} 
		Total magnetization per adatom (in $\mu_B$) 
        for different sites, calculated within GGA and GGA$+$$U$, from all-electron 
        LAPW calculations.  Site-specific values of $U$ are listed 
        in eV. }
    \begin{ruledtabular}
        \begin{tabular}{ccccc}
            Adatom & Binding site &  $m_{\text{GGA}} $ & $m_{\text{GGA}+U} $ & $U$ \\
            \hline \\[-.80em]
            Ni& O site & $0.00$ & $0.16$ & 4.6\\
            Ni& hollow site & $1.04$ & $1.06$ & 6.0 \\
            Ni& Mg site & $1.40$ & $1.66$ & 5.0\\
            Co& O site & $2.68$ & $2.78$ & 6.9\\
        \end{tabular}
    \end{ruledtabular}
\end{table}

Atomic Ni has a ground state configuration of $3d^84s^2$ ($^3F_4$) with the
next highest state $3d^94s^1$, belonging to a different multiplet ($^3D$),
only $0.025$ eV away.\cite{NIST} The average energies of these multiplets,
however, have $3d^94s^1$ $0.030$ eV lower in energy.\cite{NIST}
The proximity of these atomic states sets the stage for a
competition between chemical bonding and magnetism.\cite{markovits2001}

Our GGA calculations give a $3d^94s^1$ ground-state
configuration for atomic Ni, which underrepresents the $s$ occupation compared
to experiment, while inclusion of a local Coulomb interaction ($U=6.0~eV$)
yields a $3d^84s^2$ configuration.  In both cases, the Ni moment is
calculated to be 2.0 $\mu_B$.

The calculated magnetic moments for Ni on MgO/Ag are listed in Table \ref{tab:magmom}.  
The Ni spin moment is reduced from the atomic value on
all sites, though by varying amounts.  When Ni is on the O site, the magnetic 
moment is calculated to be zero within GGA, but it increases
slightly upon inclusion of $U$.  (When the structure is relaxed
within GGA$+$$U$ and the Ni-O bond length increases by 8\%, the moment 
increases to about 0.3 $\mu_B$.) 
On the hollow site, the magnetic moment of about 1 $\mu_B$ is
insensitive to $U$, while the larger moment of about 1.5 $\mu_B$ on the
Mg site increases with $U$. 

\begin{table}[tb]
    \caption{\label{tab:bader} 
        Net charge of Ni atom and its nearest-neighbor O and Mg atoms 
        according to Bader analysis.  Ag/MgO is the bare slab without adatoms. 
        These results were obtained from all-electron LAPW calculations.} 
    \begin{ruledtabular}
        \begin{tabular}{lcccccc}
            &  \multicolumn{3}{c}{GGA}  & \multicolumn{3}{c}{GGA$+$$U$} \\
            & Ni & $\text{O}_{nn}$ & $\text{Mg}_{nn}$ &Ni & $\text{O}_{nn}$ & $\text{Mg}_{nn}$ \\
            \hline \\[-.80em]
            O site & $-0.17$  & $-1.50$ & $+1.72$ 
            & $-0.13$ & $-1.52$ & $+1.71$\\
            hollow site & $+0.38$ & $-1.51$ & $+1.71$  
            & $+0.43$ & $-1.54$ & $+1.71$ \\
            Mg site &  $-0.26$ & $-1.62$ & $+1.68$  
            &  $-0.40$ & $-1.61$  & $+1.69$  \\
            MgO/Ag  &  & $-1.64$ & $+1.71$ \\
        \end{tabular}
    \end{ruledtabular}
\end{table}

In Table \ref{tab:bader}, we present the results of a Bader charge analysis\cite{bader} (space filling)
for the adatom and its nearest neighbor Mg and O atoms, as well as for a 
bare slab of MgO$/$Ag. Because this analysis includes the interstitial electrons, these results can provide insight on the nature of
the bonding between the adatom and the surface. On the O site, the Ni adatom gains electrons from its
O neighbor (which is less negatively charged than it would be on
bare MgO/Ag). In contrast, on the hollow site, Ni loses electrons to the 
underlying Ag substrate.  These charge transfer results are relatively 
insensitive to local Coulomb interactions.   

\begin{figure*}[]
        \begin{center}
            \includegraphics[width=\textwidth]{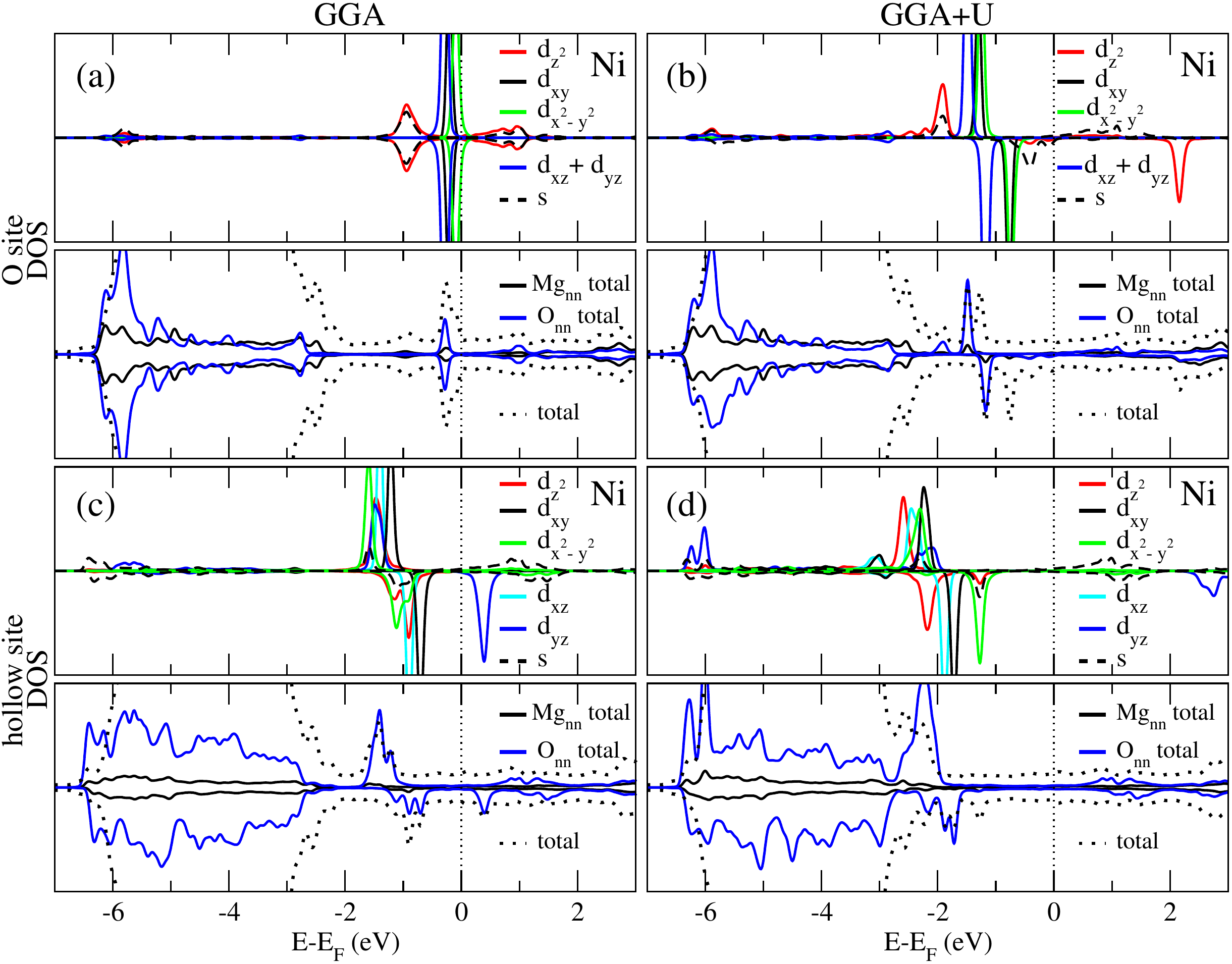}
        \end{center}
	\caption{(Color online) Densities of states  from inside the muffin-tins of 
        all-electron LAPW calculations. (a) GGA results for Ni on O site. (b)
        GGA$+$$U$ results for Ni on O site. (c) GGA results for Ni on hollow site.
        (d) GGA$+$$U$ results for Ni on hollow site. In all graphs, 
        majority spin is plotted on the positive ordinate and minority spin on the 
        negative ordinate. Total DOS has been scaled down significantly, and includes 
        contributions from the interstitial region. The relative scales of 
        nearest-neighbor Mg and O, as well as Ni $s$ and $d$ DOS have been adjusted
        to enhance viewability.}  
    \label{fig:DOS}
\end{figure*}

To understand the site dependence of the Ni moment, 
we examine the  bonding and electronic density of states, starting  
with Ni on the O site within GGA.  While the transfer of electrons  
from the substrate to a $3d^94s^1$ configuration can only 
reduce the spin moment, the gain of $\sim0.2$
is not enough to account for the full quenching 
of the moment.  
The fact that most of the electrons are transferred from 
the neighboring O atom, and the fact that Ni adsorption causes the O atom to 
displace upwards out of the MgO ML (Table \ref{tab:structure}), 
suggest the formation of a covalent  bond. 
Indeed, previous studies of Ni on MgO discuss a bonding mechanism involving  
Ni hybrid orbitals of $4s$ and $d_{z^2}$ 
character.\cite{yudanov, neyman1997,markovits2001,lopez,fernandez} 
One of the two orthogonal orbitals, which we denote as $4s$$+$$d_{z^2}$,  
is oriented perpendicular to the surface and interacts strongly with
O $p_z$  orbitals, forming a bonding combination (mostly  O $p_z$) 
about 6 eV below the Fermi level ($E_F$) and an antibonding combination (mostly
Ni $4s$$+$$d_{z^2}$)  about 1 eV above the Fermi level. 
This can be seen in the electronic density of states plotted 
in Fig. \ref{fig:DOS}(a).
The other hybrid orbital, denoted  $4s$$-$$d_{z^2}$, is concentrated in the 
plane parallel to the surface, and hence interacts much more weakly
with the surface. This planar hybrid orbital lies just below the other 
Ni $d$ states, about 1 eV below the Fermi level.  
The covalent interaction between Ni and O pushes the antibonding 
orbital high enough to be unoccupied in both spin channels,
so the other $d$ and $s$-$d$ hybrid orbitals are 
fully occupied, yielding a zero net spin. 

\begin{figure*}
	\begin{center}
		\subfigure[~O site]{\label{subfig:o}\includegraphics[width=8.5cm]{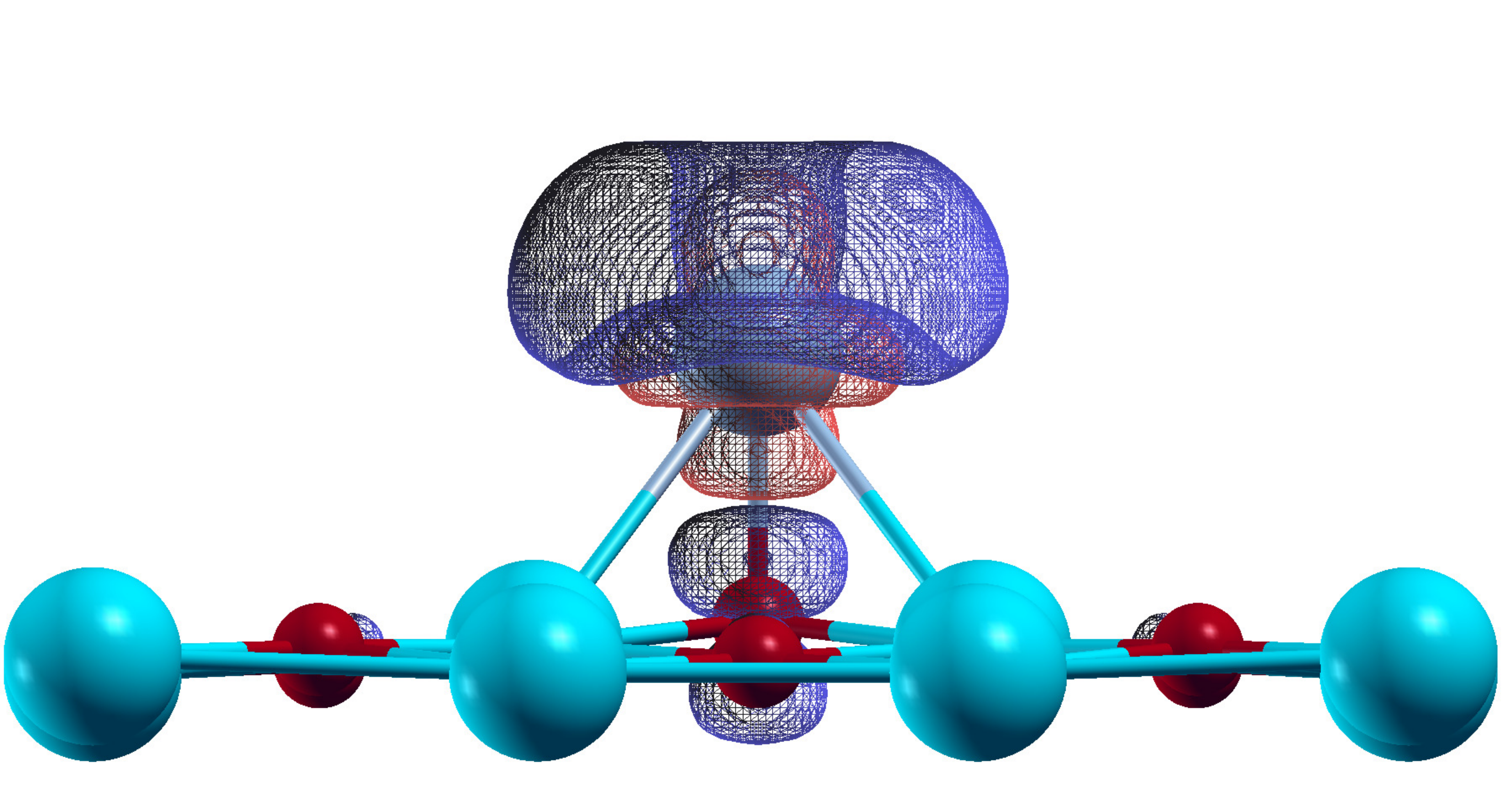}}\hspace{.2cm}\subfigure[~hollow site]{\label{subfig:h}\includegraphics[width=8.5cm]{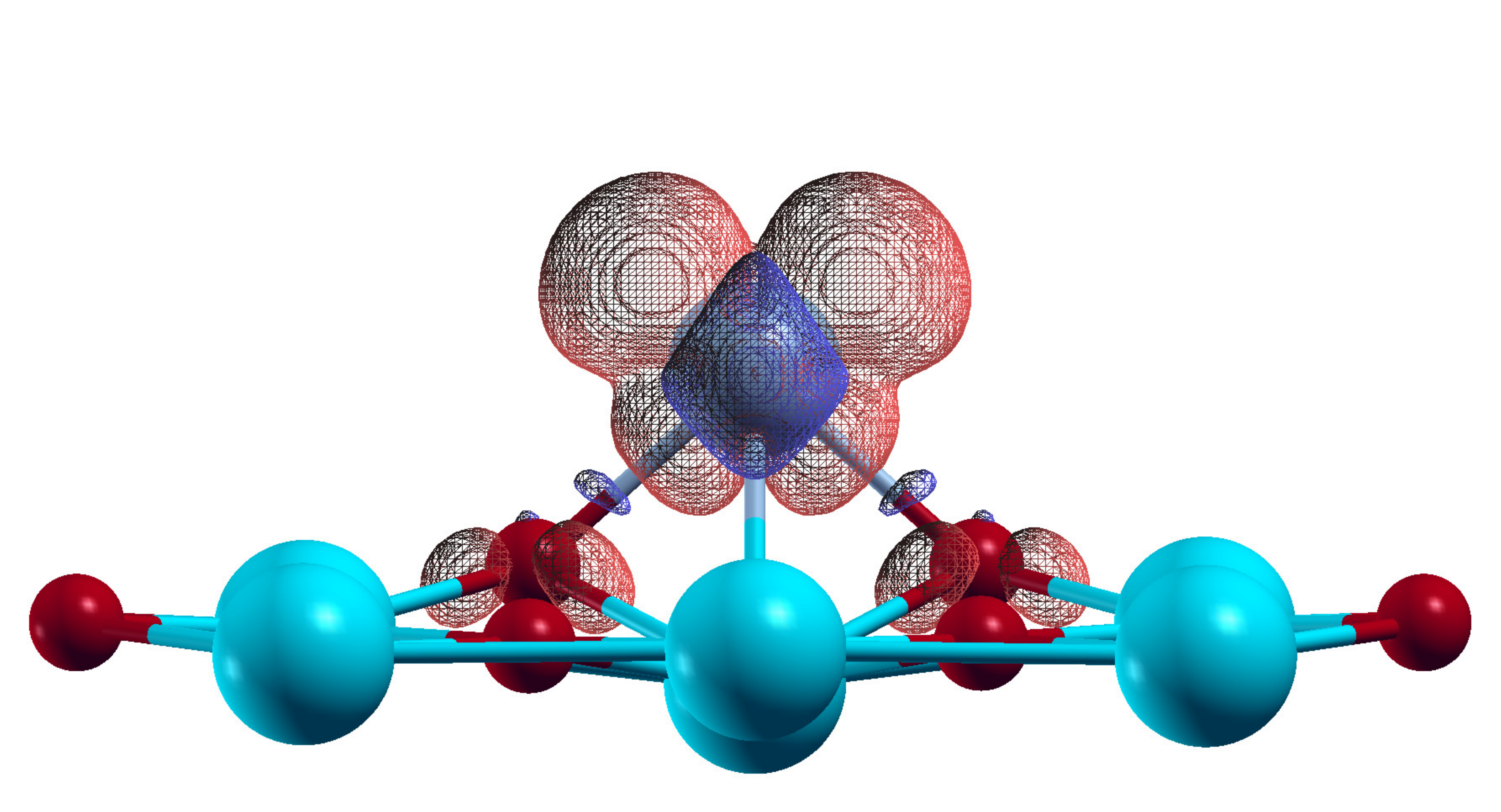}}
	\end{center}
	\caption{(Color online) Spin density isosurfaces of Ni adsorbed on MgO/Ag, from GGA$+$$U$ all-electron LAPW calculations.
        Red and blue surfaces correspond to majority and minority
        spin densities of  0.00175 e/a.u.$^3$, respectively. (a) O site: the majority spin $d_{z^2}$ is screened by the antibonding Ni $s$ - O $p$ orbital. (b) hollow site: the antibonding Ni $d_{yz}$ - O $p$ orbital is spin polarized. The y axis is 
        parallel to the line joining the two O atoms that neighbor the Ni. 
        These two O atoms are vertically displaced out of the MgO plane. O atoms are represented by small red spheres.
        \label{fig:spin}}
\end{figure*}

\indent When the orbital-dependent Coulomb interaction $U$ is
included within GGA$+$$U$, the density of state changes
significantly since the $d$ and $s$ states are
affected differently. The hybridization  between Ni $d_{z^2}$ and $s$ orbitals 
weakens, and the interaction between these orbitals and O $p$ states
is affected.  The partial density of states plots in Fig. \ref{fig:DOS}(b) show
the majority and minority spin $d_{z^2}$ orbitals split by about 4 eV 
($\sim U$). The antibonding Ni - O  orbital, now mostly composed of Ni $s$ states, 
has a small spin splitting of the opposite polarity: the spin-down 
states lie just below $E_F$ and the spin-up ones lie just above $E_F$.
Hence even though $U$ induces a moment in the $d$ shell within the 
Ni muffin tin, this moment is screened by the oppositely polarized
antibonding orbital formed from Ni $4s$  and O $2p_z$ orbitals.  
This is illustrated in Fig. \ref{fig:spin}\subref{subfig:o},
which shows a majority-spin (red) isosurface with $d_{z^2}$ character on the Ni
atom, surrounded by the antibonding Ni $s$ -  O $p_z$ minority-spin  (blue)
isosurface (corresponding to minority spin states near $\sim -0.5~eV$ in Fig. \ref{fig:DOS}(b)).  The total magnetization of the system remains close to zero.

The situation differs for the hollow site. 
The presence of the Ni at the hollow site disrupts the MgO bond network,  
and the two O atoms that neighbor the Ni atom are displaced out of the surface 
by nearly 0.5 \AA~ to form  bonds with the Ni, as is evident in  
Fig. \ref{fig:spin}\subref{subfig:h}. The resulting polar covalent 
Ni-O bond length is only about 6\% larger than the corresponding O-site bond
(Table \ref{tab:structure}). 
The hollow site has $\text{C}_{2v}$ symmetry, meaning that the
$d_{xz},d_{yz}$ degeneracy is broken. The lobes of the $d_{yz}$ orbital 
point towards the O neighbors and interact with O $p_y$ orbitals. 
However, since $d_{yz}$ transforms differently than $4s$, 
symmetry considerations prevent the $d_{yz}$ orbital from forming diffuse 
hybrids with $4s$ analogous to the $4s$$\pm$$d_{z^2}$ orbitals on the O site
that enhance the orbital overlap between the adatom and surface atoms.
While symmetry permits hybridization between Ni $s$ and $d_{x^2-y^2}$ orbitals,
the hybrid that interacts strongly with O $p$ states 
has mostly Ni $s$ character, while the orthogonal hybrid orbital is mostly 
Ni $d_{x^2-y^2}$. Hence the Ni $d$ orbitals all retain a higher degree
of localization than they do on the O site. 
Within GGA, it becomes favorable for the antibonding Ni $d_{yz}$ -  O $p$ orbitals to become spin 
polarized and this largely accounts for the total moment of about 1 $\mu_B$. 
This can be seen in Fig. \ref{fig:DOS}(c).

Because the effect of $U$ in the hollow site case is to increase the 
splitting between the $d_{yz}$ majority and minority spin states, the 
occupation of $s$ and $d$ states and the total moment are not significantly 
impacted, as can be seen in Fig. \ref {fig:DOS}(d).
Figure \ref{fig:spin}\subref{subfig:h} shows the spin density 
for this case. The majority-spin isosurface in red has 
antibonding Ni $d_{yz}$ - O $p_y$  character. While it is hard to see 
from the angle shown, the minority-spin isosurface in blue has the character 
of a $4s$$-$$d_{x^2-y^2}$ hybrid. 

We return now to the question of why the local Coulomb 
interaction stabilizes the hollow site relative to the O site.
When Ni adsorbs to the O site it forms a strong covalent bond 
with O through diffuse $sd$ hybrids that overlap significantly with 
O $p$ orbitals. Since the orbital-dependent $U$ disrupts the
$sd$ hybridization, it weakens the binding of the Ni to the O. 
On the hollow site, geometric and symmetry considerations 
prevent the formation of diffuse $sd$ hybrids that overlap strongly 
with O, so the impact of $U$, both in terms of binding energy
and spin moment, is smaller.  As a result, increasing the  
strength of the local Coulomb interaction decreases the stability 
of the O site relative to the hollow site. 

The presence of magnetization in adatoms depends largely on the 
relative strengths of the covalent interaction with surface atoms
and the intra-atomic exchange coupling for the adatom.  In a 
previous study,\cite{cobalt} 
our calculations found that Co adsorbs to the O site of the MgO/Ag surface. 
We also found that, in contrast to Ni, it maintains a magnetization close 
to the atomic value (Table \ref{tab:magmom}).  This can be explained by 
differences in the Ni-O and Co-O interactions. Unlike Ni, Co remains charge 
neutral when adsorbed, according to a Bader analysis, and the 
neighboring O and Mg atoms displace downward rather than being attracted to 
the adatom (Table \ref{tab:structure}). 
Since the $s$-$d$ transition energy is larger in Co than Ni, the
hybridization between $s$ and $d$ orbitals is reduced, leading to 
less interatomic overlap and weaker covalent interactions.  The $d$
orbitals remain localized enough to support magnetism.

\section{\label{sec:conclusions}Conclusions}
Using first-principles calculations, we have found a strong site dependence of the electronic and magnetic properties of Ni on MgO/Ag. In particular, the magnetic moment varies from S $ = 0$ to nearly 1 as Ni is moved away from surface O. This is in stark contrast with the situation of Co on MgO/Ag, in which the O site facilitates the preservation of Co spin and orbital moments on the surface. 
\\ \indent Furthermore, we see that the energy
surface for adsorption of Ni on MgO/Ag is very different from 
that for Ni on MgO. On MgO/Ag, the O site and hollow site are 
competitive in energy,  and it is not clear from theory alone which site 
will be realized in experiments.  However, the properties of the Ni adatom
on the two sites differ significantly, not just in the magnetic moment, but also in
the charge transfer and bonding.

The energetic proximity of the hollow and O sites raises the interesting
possibility of being able to select one site or the other, and tune the magnetic moment by modification of the 
substrate through, for example, application of strain or introduction of 
defects.  It would also be interesting to compare bilayer and monolayer MgO
on Ag experimentally.
The bilayer would be much more rigid than the monolayer, and hence we would 
expect the O site to be clearly favored for Ni adsorption.  

\begin{acknowledgments}
The authors thank Shruba Gangopadhyay, Susanne Baumann and Andreas Heinrich for fruitful discussions. This work was supported in part by NSF Grants DMR-1006605 \& EFRI-143307. BAJ acknowledges the Aspen Center for Physics and the NSF Grant PHY-1066293 for hospitality during the writing of this paper.
\end{acknowledgments}

\end{document}